\documentclass[aps,prb,groupedaddress,preprint]{revtex4-1}
\usepackage{graphicx}
\usepackage{helvet}
\usepackage{amsmath,amssymb}
\usepackage{bm}
\usepackage{mathrsfs}

\begin{document}

\title{A spin-gapped Mott insulator with the dimeric arrangement of twisted molecules Zn(tmdt)$_{2}$}
\author{Rina Takagi$^{1}$, Hiro Gangi$^{1}$, Kazuya\ Miyagawa$^{1}$, Biao\ Zhou$^{2}$, Akiko Kobayashi$^{2}$, and Kazushi Kanoda$^{1}$}
\affiliation{$^{1}$Department of Applied Physics, University of Tokyo, Bunkyo-ku, Tokyo, 113-8656, Japan \\
$^{2}$Department of Chemistry, College of Humanities and Sciences, Nihon University, Setagaya-ku, Tokyo, 156-8550, Japan}

\begin{abstract}
$^{13}$C nuclear magnetic resonance measurements were performed for a single-component molecular material Zn(tmdt)$_{2}$, in which tmdt's form an arrangement similar to the so-called ${\kappa}$-type molecular packing in quasi-two-dimensional Mott insulators and superconductors.
Detailed analysis of the powder spectra uncovered local spin susceptibility in the tmdt ${\pi}$ orbitals. 
The obtained shift and relaxation rate revealed the singlet-triplet excitations of the ${\pi}$ spins, indicating that Zn(tmdt)$_{2}$ is a spin-gapped Mott insulator with exceptionally large electron correlations compared to conventional molecular Mott systems.
\end{abstract}

\maketitle

A new type of molecular conductors $M$(tmdt)$_{2}$ ($M =$ Ni, Au, Pt, Cu, and Pd) are of current interest not only because it is the first molecular metal composed of a single molecular species \cite{1} but also affords novel electronic states owing to the multi-orbital nature, which conventional charge-transfer salts do not possess. 
Furthermore, the systems of $M$(tmdt)$_{2}$ are expected to host highly correlated electrons compared with conventional molecular materials with dimeric molecular structures called $2:1$ compounds, where a dimer playing as one lattice site accommodates one carrier and thus the on-site Coulomb repulsion $U$ is reduced by the spatial extension of the dimer orbital.
In case that one molecule accommodates one carrier as in some of the systems $M$(tmdt)$_{2}$, the $U$ is not reduced by the dimerization.

The $M$(tmdt)$_{2}$ consists of a single molecular species in which a transition-metal ion, $M$, is coordinated by molecular ligands, tmdt, from both sides \cite{2,3}.
The $p{\pi}$ molecular orbitals extended over the tmdt ligand lie near the Fermi level, whereas the energy level of the $dp{\sigma}$ molecular orbital located around $M$ strongly depends on $M$.\cite{4,5,6,7}
For $M =$ Cu, the $dp{\sigma}$ orbital lies close to the $p{\pi}$ orbitals, leading to the formation of a $dp{\sigma}$-$p{\pi}$ multi-orbital system \cite{5,7}; indeed, NMR studies of Cu(tmdt)$_{2}$ found that the $dp{\sigma}$ and $p{\pi}$ orbitals host antiferromagnetic and spin-gapped Mott subsystems, respectively \cite{8,9}.
For $M =$ Ni, Pt and Zn, the $dp{\sigma}$ orbitals lie away (upward in the former two and downward in the latter) from the two $p{\pi}$ orbitals located near the Fermi level \cite{4,10}. 
In these systems, two $p{\pi}$ orbitals in $M$(tmdt)$_{2}$ accommodate two electrons, likely giving $p{\pi}$-band semimetals; the $ab$-initio band structure calculation predicts small Fermi pockets \cite{4,10,11}.
As expected, Ni(tmdt)$_{2}$ and Pt(tmdt)$_{2}$ are metals \cite{1,10,12} and appreciable electron correlation was revealed by $^{13}$C nuclear magnetic resonance (NMR) studies \cite{13}.

It is to be noticed, however, that Zn(tmdt)$_{2}$ is an insulator\cite{14}; the typical resistivity data are shown in Fig. 1(d) although only two-terminal measurements are available to date because of tiny crystals.
The high-temperature part of the Arrhenius plot shown in the inset gives an estimate of the activation energy of 0.091 eV, assuming that the low-temperature part is affected by the impurity or defect levels formed in the gap, as usual in semiconductors.
The magnetic susceptibility rapidly decreases with lowering temperature and nearly vanishes below approximately 200 K,\cite{14} indicating a nonmagnetic insulating ground state, which does not accord with the band-structure calculation.
Its structure is of additional interest. While two tmdt's in $M$(tmdt)$_{2}$ are in the same plane for $M =$ Ni, Pt, and Cu, they are twisted with a dihedral angle of $84.9^{\circ}$ in Zn(tmdt)$_{2}$,\cite{14} as shown in Fig. 1(a).
In the solid, two tmdt ligands are arranged face-to-face and they form a checker-board pattern in two-dimensional layers (Figs. 1(b) and (c)).
This type of molecular arrangement is similar to the ``${\kappa}$-type configuration'', which is familiar in the $2:1$ charge-transfer salts represented by ${\kappa}$-(ET)$_{2}X$, including dimer-Mott insulators and over-10-K superconductors \cite{16}.
Most remarkably, Zn(tmdt)$_{2}$ accommodates one electron per tmdt in the ${\kappa}$-type configuration in contrast to one hole accommodated by two ET's in ${\kappa}$-(ET)$_{2}X$; so to speak, Zn(tmdt)$_{2}$ is an equivalent to yet unknown ${\kappa}$-type $1:1$ charge transfer salt.

In the present study, we have investigated in-depth electronic state of Zn(tmdt)$_{2}$ by using $^{13}$C NMR technique. 
Our finding is that Zn(tmdt)$_{2}$ is a spin-gapped Mott insulator with extraordinarily strong electron correlation among organic materials.

For $^{13}$C NMR, the $^{13}$C isotope is selectively enriched at the carbon site in tmdt as shown in Fig. 1(a).
The solid samples of Zn(tmdt)$_{2}$ were synthesized by the method as described in Appendix.
$^{13}$C NMR spectra and nuclear spin-lattice relaxation rate, $T_{1}^{-1}$, were measured for an assembly of fine polycrystals of $^{13}$C-enriched Zn(tmdt)$_{2}$ under a magnetic field of 8.00 Tesla.
The spectra were obtained by the fast Fourier transformation of echo signals observed after the spin echo pulse sequences of $({\pi}/2)_{x}$-$({\pi})_{x}$.
The ${\pi}/2$ pulse width was 1.8-3.4 ${\mu}$s, which was sufficient to cover the whole spectral frequency.
As the origin of the NMR shift, we referred to the $^{13}$C NMR line of TMS (tetramethylsilane).
$^{13}$C nuclear spin-lattice relaxation rate $T_{1}^{-1}$ was obtained by the standard saturation-recovery method.

$^{13}$C NMR spectra for Zn(tmdt)$_{2}$ are shown in Fig. 2(a).
Around room temperature, the line shape is similar to powder patterns of the Knight shift with uniaxial symmetry. 
This feature is consistent with the $p{\pi}$-orbital character of the conduction band because the spins on the $p{\pi}$ orbitals generate anisotropic hyperfine fields at $^{13}$C sites mainly through the on-site $2p_{z}$ dipolar coupling.
As temperature is decreased, the line shape varies such that triaxial symmetry becomes visualized, reflecting the temperature dependence of the three principal values of shift tensor, ${\delta}_{xx}$, ${\delta}_{yy}$ and ${\delta}_{zz}$.
It is noted that there is no indication of line broadening arising from the generation of internal fields due to magnetic ordering, which demonstrates the nonmagnetic ground state for Zn(tmdt)$_{2}$.
The shift under a field described in spherical coordinates, (${\theta}$, ${\phi}$), is given by ${\delta}({\theta}, {\phi}) = {\delta}_{xx}{\sin}^2{\theta}{\cos}^2{\phi} + {\delta}_{yy}{\sin}^2{\theta}{\sin}^2{\phi} + {\delta}_{zz}{\cos}^2{\theta}$.
The distribution function of ${\delta}$(${\theta}$, ${\phi}$) against random spherical coordinates, $f({\delta})$, should give the spectral shape of a powder sample.
In actuality, inevitable inhomogeneity causes additional broadening of the spectra.
Thus, in the analysis, we incorporated the inhomogeneous broadening by convoluting $f({\delta})$ with a Lorentzian function of the following form
\begin{eqnarray}
F({\delta}) &=& {\int} f({\nu}) \frac{w}{({\nu}-{\delta})^{2}+w^{2}} d{\nu} \nonumber\\
&=& {\iint} \frac{w}{({\delta}_{xx}{\sin}^{2}{\theta} {\cos}^{2}{\phi} +{\delta}_{yy}{\sin}^{2}{\theta}{\sin}^{2}{\phi}+{\delta}_{zz}{\cos}^{2}{\theta}-{\delta})^{2}+w^{2}} {\sin}{\theta} d{\theta}d{\phi},
\end{eqnarray}
where $w$ characterizes the inhomogeneous width. 
The fitting curves are shown with red lines in Fig. 2(a); to reduce the number of fitting parameters, we first determined the center of gravity of each spectrum, namely the isotropic component of shift, ${\delta}_{\rm iso}$, by its first moment and proceeded to the spectral fitting under the constraint of ${\delta}_{\rm iso} = ({\delta}_{xx}+{\delta}_{yy}+{\delta}_{zz})/3$.
The values of ${\delta}_{xx}$, ${\delta}_{yy}$ and ${\delta}_{zz}$, to best fit the spectra are shown as a function of temperature in Fig. 2(b).

The shift tensor, \textrm{\boldmath ${\delta}$}, arises from the temperature-independent chemical shift and temperature-dependent Knight shift tensors, \textrm{\boldmath ${\sigma}$} and \textrm{\boldmath $K$}($T$), respectively; namely, \textrm{\boldmath ${\delta}$} = \textrm{\boldmath ${\sigma}$} + \textrm{\boldmath $K$}($T$), where \textrm{\boldmath ${\sigma}$} depends on the local chemical structure around the nuclear site while \textrm{\boldmath $K$}($T$) reflects the spin susceptibility. 
The \textrm{\boldmath ${\delta}$} is explicitly expressed as
\begin{equation}
\begin{pmatrix}
\delta_{xx} &0 &0\\
0 &{\delta}_{yy} &0\\
0 &0 &{\delta}_{zz}
\end{pmatrix}
= {\sigma}_{\rm iso} + 
\begin{pmatrix}
{\sigma}_{1} &0 &0\\
0 &{\sigma}_{2} &0\\
0 &0 &{\sigma}_{3}
\end{pmatrix}
+ K_{\rm iso}(T) + K_{\rm aniso}(T)
\begin{pmatrix}
-1 &0 &0\\
0 &-1 &0\\
0 &0 &2
\end{pmatrix}
,
\end{equation}
where $x$ and $y$ axes are in the tmdt plane and $z$ axis is perpendicular to them as shown in Fig. 1(a). 
The first and second (third and fourth) terms express the isotropic and anisotropic components of chemical shift (Knight shift) tensor, respectively; ${\sigma}_{1}+{\sigma}_{2}+{\sigma}_{3}=0$ and the form of the fourth term assumes the uniaxial symmetry of the dipolar hyperfine tensor stemming from the on-site $2p_{z}$ spin.
At low temperatures, where the spin susceptibility vanishes in Zn(tmdt)$_{2}$, \textrm{\boldmath ${\delta}$} should be equal to the chemical shift, \textrm{\boldmath ${\sigma}$}.
The low-temperature saturated values in Fig. 2(b) yield the chemical shift values of ${\sigma}_{\rm iso} = 126$, ${\sigma}_{1} = 47.0$, ${\sigma}_{2} = -4.4$, and ${\sigma}_{3} = -42.8$ (in the unit of ppm).

The temperature dependences of the isotropic component, ${\delta}_{\rm iso} (= {\sigma}_{\rm iso} + K_{\rm iso})$, and the anisotropic components, $^{1}{\delta}_{\rm aniso} (={\sigma}_{1} - K_{\rm aniso})$, $^{2}{\delta}_{\rm aniso}$ ($={\sigma}_{2} - K_{\rm aniso}$) and $^{3}{\delta}_{\rm aniso}$ ($= {\sigma}_{3} + 2K_{\rm aniso}$), are shown in Figs. 3(a) and (b).
The temperature dependent parts come from the Knight shift components (the third and fourth terms in Eq. (2)). 
As seen in Fig. 3(a), the $K_{\rm iso}$ nearly follows the temperature dependence of the spin susceptibility, ${\chi}$, which rapidly decreases with temperature and reaches zero below 100 K (Fig. 3(c)).
This is consistent with only a single kind of orbitals responsible for the electronic bands, namely the $p{\pi}$ orbitals.
Indeed, the $K_{\rm iso}$-${\chi}$ plot yields a good linearity (Fig. 3(d)) as expected. 
Fitting the data for 200-300 K with a linear function yields the isotropic hyperfine coupling constant, $a_{\rm iso} = 3.2$ kOe/(${\mu}_{\rm B}$ tmdt) in the form of $K_{\rm iso} = a_{\rm iso}{\chi}$.

As seen in Fig. 3(b), the anisotropic components, $^{i}{\delta}_{\rm aniso}$, also show the temperature dependence similar to that of the spin susceptibility, ${\chi}$.
Like $K_{\rm iso}$, $K_{\rm aniso}$ should be expressed as $^{i}K_{\rm aniso} = -a_{\rm aniso} {\chi}$ ($i=1, 2$) and $^{3}K_{\rm aniso} = 2a_{\rm aniso} {\chi}$, where $a_{\rm aniso}$ is the anisotropic hyperfine coupling constant.
As expected, two of the three branches in Fig. 3(b) shows approximately the same temperature dependence in sign and magnitude, while the other branch varies opposite in sign and twice in magnitude.
These characteristics are totally as expected in the form of $^{i}K_{\rm aniso}$.
The $^{i}K_{\rm aniso}$-${\chi}$ plots are displayed in Fig. 3(e). 
The fitting gave $a_{\rm aniso} = 2.1$ kOe/(${\mu}_{\rm B}$ tmdt), which is the average of the $a_{\rm aniso}$ values given by the slopes of the $^{i}K_{\rm aniso}$-${\chi}$ plots in Fig. 3(e).

The hyperfine coupling constants, $a_{\rm iso}$ and $a_{\rm aniso}$, and the ratio, $a_{\rm aniso}/a_{\rm iso}$, for $M$(tmdt)$_{2}$ ($M=$ Zn, Ni, Pt and Cu) are listed in TABLE I. 
We note that both of the $a_{\rm iso}$ and $a_{\rm aniso}$ values are appreciably smaller for $M=$ Zn than the others.
The values of $a_{\rm iso}$ and $a_{\rm aniso}$ measure the local spin density around the $^{13}$C site through the Fermi contact (and/or core polarization) and dipolar interactions, respectively \cite{NMR}.
Thus, it is suggested that the electron population of the $p_{z}$ orbital at the $^{13}$C site, which has a major contribution to the hyperfine coupling, is particularly small in Zn(tmdt)$_{2}$ compared with the others. 
In other words, the spatial profile of the highest occupied molecular orbital (HOMO) of tmdt in Zn(tmdt)$_{2}$ is modified from those in other $M$(tmdt)$_{2}$'s. 
This is reflected on the exceptionally small ratio of $a_{\rm aniso}/a_{\rm iso} = 0.66$ for Zn(tmdt)$_{2}$, compared with the ratios of $0.8{\pm}0.02$ for others \cite{9,13}.
A possible origin is the effect of a twist in the molecular structure. 
The effect of the hybridization of $d$ orbital to $p{\pi}$ orbital is also conceivable. 
It is pointed out that the HOMO of Ni(tmdt)$_{2}$ and the corresponding orbitals of the other $M$(tmdt)$_{2}$ with planar molecular structure have a small but finite $d$-orbital hybridization \cite{4,5,6,7}, which should modify the $p{\pi}$ orbital profile.
Such $d$-orbital hybridization in the HOMO is, however, absent in Zn(tmdt)$_{2}$,\cite{4} possibly making difference between Zn(tmdt)$_{2}$ and the other compounds.

$^{13}$C nuclear spin-lattice relaxation rate $T_{1}^{-1}$ is plotted in Fig. 4(a). 
The recovery of nuclear magnetization in the present powder sample was not an exponential function of time because the local field at the $^{13}$C sites is distributed from grain to grain, depending on the geometry of the field direction against the crystal axes owing to the anisotropic hyperfine coupling. 
Then, we obtained $T_{1}^{-1}$ by fitting the relaxation curves to the stretched exponential function, $M(t)=M_{0}\left\{ 1- \exp[-(t/T_{1})^{\beta}]\right\}$, where $M(t)$ is the nuclear magnetization at a time, $t$, after its saturation caused by the so-called rf comb pulse. 
The exponent ${\beta}$ characterizes the degree of distribution in $T_{1}$.
Near room temperature ($T > 220$ K), however, an additional exponential term is required to fit the relaxation curve over the entire time scale; thus, we adopted the form of $M(t)=M_{0}\left\{ 1-C \exp[-(t/T_{1})^{\beta}]-(1-C) \exp[-(t/ ^{\rm L}T_{1})]\right\}$ as a fitting function. 
The second $^{\rm L}T_{1}$ term has a contribution of only 10 ${\%}$ to the nuclear magnetization and thus likely originates from an unknown impurity phase. 
For $T > 220$ K, the values of $T_{1}^{-1}$ (blue squares in Fig. 4(a)) is much larger than $^{\rm L}T_{1}^{-1}$ (red triangles in Fig. 4(a)); thus, they could be separated. 
Below 200 K, the two values approach each other, making them inseparable. However, considering only 10 ${\%}$ contribution of the second phase, the influence of the impurity phase to the $T_{1}^{-1}$ values obtained in the present analysis is not significant even for $T < 200$ K. 
The exponent ${\beta}$ is about 0.8 for $T > 220$ K; however, it gradually decreases with lowering temperature and reaches 0.5-0.6 below 100 K (Fig. 4(b)), indicating an additional distribution in $T_{1}$ at low temperatures.

The $T_{1}^{-1}$ rapidly decreases by three orders of magnitude on cooling from room temperature to 10 K, which indicates paramagnetic spins thermally activated over an energy gap.
Regarding the origin of the gap, two cases are conceivable. 
One case is a band insulator, for which the band comprising the bonding orbitals in the tmdt dimers is fully occupied and has an energy gap to an unoccupied antibonding band. 
The other case is a dimerized Mott insulator, where one $p{\pi}$ electron is localized on a tmdt because of Coulomb interactions and the spins form singlets due to the dimerization. 
The former case is, however, clearly ruled out.
According to the first-principles band calculation \cite{4}, the density of states (DOS) at the Fermi level (${\epsilon}_{\rm F}$) is ${\sim}4$ electrons/eV/molecule for the $p{\pi}$-band paramagnetic metal Ni(tmdt)$_{2}$, while the DOS averaged over ${\epsilon}_{\rm F}{\pm}300$ K is ${\sim}1$ electrons/eV/molecule for Zn(tmdt)$_{2}$.
Given that Zn(tmdt)$_{2}$ is a band insulator, $(T_{1}T)^{-1}$ should show a temperature variation of the Arrhenius type \cite{9}. 
In this case, the band gap turns out to be $2{\Delta} = 2100$ K (or 0.18 eV) by fitting the data by the form of $(T_{1}T)^{-1} {\propto} {\exp}(-{\Delta}/T)$ for 200-300 K, where the relaxation rate of the sample is well separated from that of the minor impurity phase (the inset of Fig. 4(c)). 
Then, the number of paramagnetic spins thermally activated over the gap of 2100 K at room temperature should be orders of magnitude smaller than that of Ni(tmdt)$_{2}$; equivalently, the $T_{1}^{-1}$ values in the former should be overwhelmed by the latter. 
However, the value of $T_{1}^{-1}$ for Zn(tmdt)$_{2}$ is 42 sec$^{-1}$ at 297 K, which is even one order of magnitude larger than that for Ni(tmdt)$_{2}$, ${\sim}$3 sec$^{-1}$.

In the case of the dimerized Mott insulator, the relaxation originates from the singlet-triplet excitations over a spin gap of ${\Delta}_{s} = 1300$ K (or 0.11 eV), obtained from the Arrhenius plot of $T_{1}^{-1}$ for 200-300 K (Fig. 4(c)). 
The hyperfine field at the $^{13}$C site from $S = 1/2$ localized $p{\pi}$ spins can be rather large compared to that from conducting $p{\pi}$ spins. 
Because the arrangement of tmdt ligands in Zn(tmdt)$_{2}$ is dimeric, a spin singlet should be formed in a dimer as shown in Fig. 4(d). 
According to the singlet-triplet excitation model, the magnetic susceptibility is expressed by
\begin{equation}	
\chi = \frac{2N_{\rm A}g^{2}{\mu}_{\rm B}^{2}}{3k_{\rm B}T} \frac{3e^{(-{\Delta}_{s}/T)}}{1+3e^{(-{\Delta}_{s}/T)}},
\end{equation}
where $N_{\rm A}$ is the Avogadro constant, $g$ is the $g$ factor, ${\mu}_{\rm B}$ is the Bohr magneton, and $k_{\rm B}$ is the Boltzmann constant. 
The experimentally obtained ${\chi}$ was well fitted with only one parameter ${\Delta}_{s}$, which was estimated at ${\Delta}_{s} = 1400$ K, as shown by the red curve in Fig. 3(c). 
The gap value is in agreement with that evaluated by the $T_{1}^{-1}$ analysis. 
The curves in Figs. 3(a) and (b) show calculations with the use of the hyperfine coupling constants obtained above and the ${\chi}$ curve represented by the red curve in Fig. 3(c), reproducing the experimental shift data.
Thus, we conclude that Zn(tmdt)$_{2}$ is a spin-gapped Mott insulator, where one tmdt accommodates one electron.

The present results revealed the $p{\pi}$ Mott insulating phase in Zn(tmdt)$_{2}$. 
Assuming the Heisenberg type of spin coupling, the exchange interaction is described as $J = 4t^{2}/U_{p{\pi}}$, where $U_{p{\pi}}$ is the Coulomb repulsive energy on one tmdt. 
Most simply, the energy gap for the singlet-triplet excitations, ${\Delta}_{s}$, corresponds to the largest exchange interaction. 
Actually, in Ref. [9], $U_{p{\pi}}$ of Cu(tmdt)$_{2}$, which was also found to host spin-gapped Mott insulating $p{\pi}$ spins, was estimated at 2.3 eV by using the values of ${\Delta}_{s} = 1550$ K and $t = 0.25$ eV (the largest transfer integral between the tmdt ligands) \cite{7}. 
Because the transfer integral in Zn(tmdt)$_{2}$ is unknown, we assume the $t$ value in Cu(tmdt)$_{2}$ to estimate $U_{p{\pi}}$ in Zn(tmdt)$_{2}$ and find $U_{p{\pi}} = 2.1$ eV by using ${\Delta}_{s} = 1350$ K (the average of the gap values determined from the relaxation rate and the magnetic susceptibility). 
For reference, the $U$ value in charge-transfer salts with dimeric molecular structures is typically ${\sim}1$ eV or less.
Thus, Zn(tmdt)$_{2}$ is addressed as a highly correlated $p{\pi}$ system, and such strong electron correlation is found common in $M$(tmdt)$_{2}$, regardless of molecular structures or $dp{\sigma}$-orbital states.
We emphasize that Zn(tmdt)$_{2}$ provides a unique example of the ``${\kappa}$-type'' arrangement of molecules, which accommodate one electron per ligand.

\section*{ACKNOWLEDGEMENTS}
We thank H. Seo for helpful discussions.
This work was supported by the JSPS Grant-in-Aids for Scientific Research (S) (Grant No. 25220709) and for Scientific Research (C) (Grant No. 17K05846), and the JSPS Fellows (Grant No. 13J03087).

\section*{Appendix}
As shown in Scheme 1, the $^{13}$C-enriched tmdt ligands were synthesized from $^{13}$CS$_{2}$ according to the literature \cite{17,18}, and the $^{13}$C-enriched Zn(tmdt)$_{2}$ was synthesized according to Scheme 2 similar to the normal Zn(tmdt)$_{2}$ reported previously \cite{14}.
The deprotection of the $^{13}$C-enriched tmdt ligand using the MeOH solution of tetramethylammonium hydroxide in dry THF at room temperature yielded (Me$_{4}$N$^{+}$)$_{2}$($^{13}$C-tmdt$^{2-}$).
After cooling to -78 $^{\circ}$C with a methanol-dry ice bath, the MeOH solution of (Me$_{4}$N)$_{2}$[ZnCl$_{4}$] was added to the system.
Then, pale yellow crystals of $^{13}$C-enriched (Me$_{4}$N)$_{2}$[Zn(tmdt)$_{2}$] were obtained.
The single crystals of $^{13}$C-enriched Zn(tmdt)$_{2}$ were grown by the electrochemical oxidation of $^{13}$C-enriched (Me$_{4}$N)$_{2}$[Zn(tmdt)$_{2}$] using $^{n}$Bu$_{4}$N${\cdot}$PF$_{6}$ as an electrolyte and acetonitrile as a solvent. 
In the electrochemical oxidation, constant current of 0.4 ${\mu}$A was applied between two platinum electrodes under an argon atmosphere during 4 weeks.

\newpage
\begin{figure}
\begin{center}
\includegraphics[width=12cm,keepaspectratio]{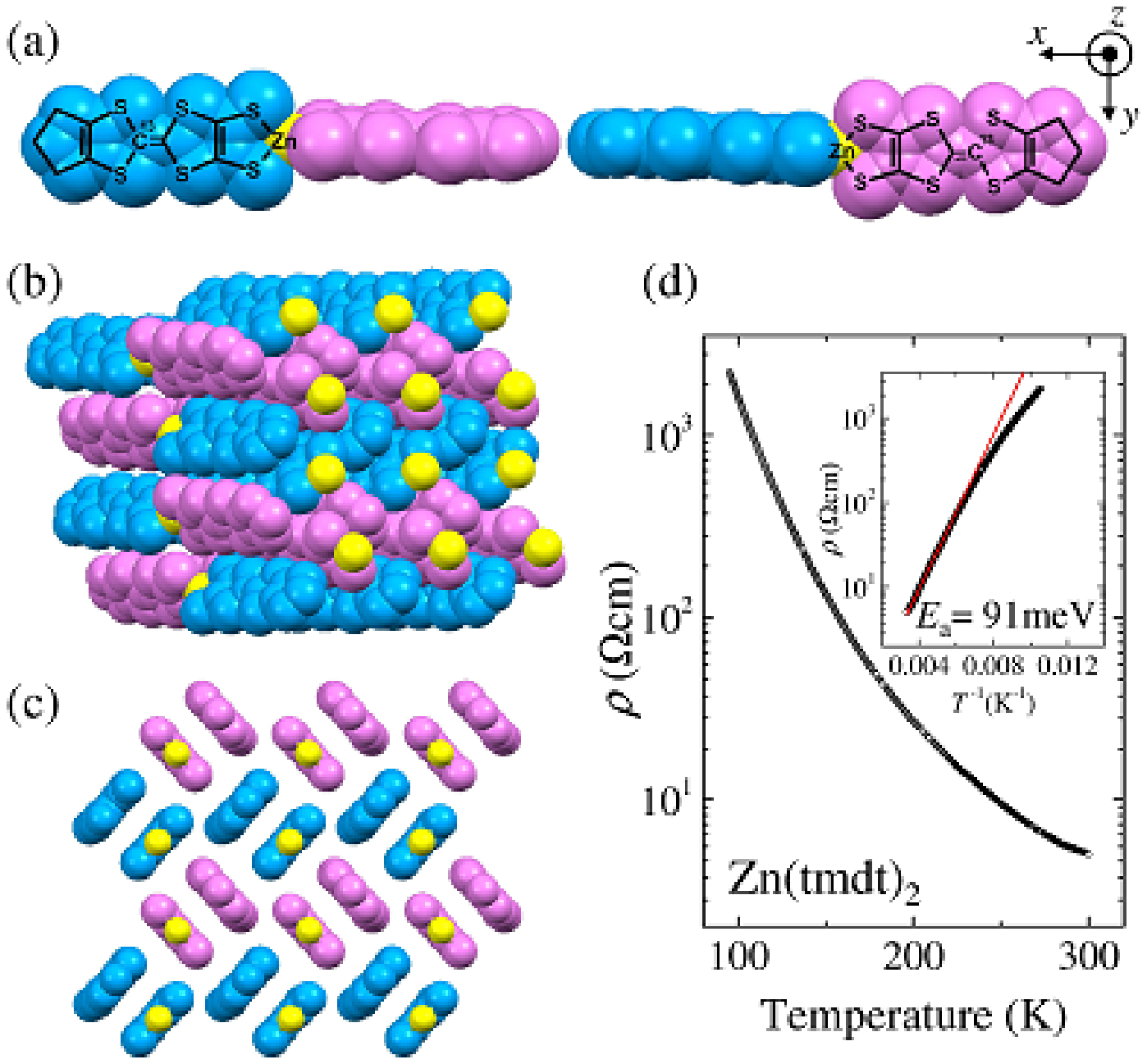}
\end{center}
\caption{(Color online) (a) Molecular structure of Zn(tmdt)$_{2}$. 
(b) Molecular arrangement in the Zn(tmdt)$_{2}$ crystal. 
(c) End-on projection of the ``${\kappa}$-type molecular arrangement'' in the tmdt layer.
(d) Temperature dependence of two-probe resistivity for Zn(tmdt)$_{2}$ measured on a single crystal of approximately 70 ${\mu}$m.
The inset shows the Arrhenius plot and an activation energy $E_{\rm a}=91$ meV was obtained from the data for 200-280 K (indicated by a red curve).}
\label{Fig1}
\end{figure}

\begin{figure}
\begin{center}
\includegraphics[width=12cm,keepaspectratio]{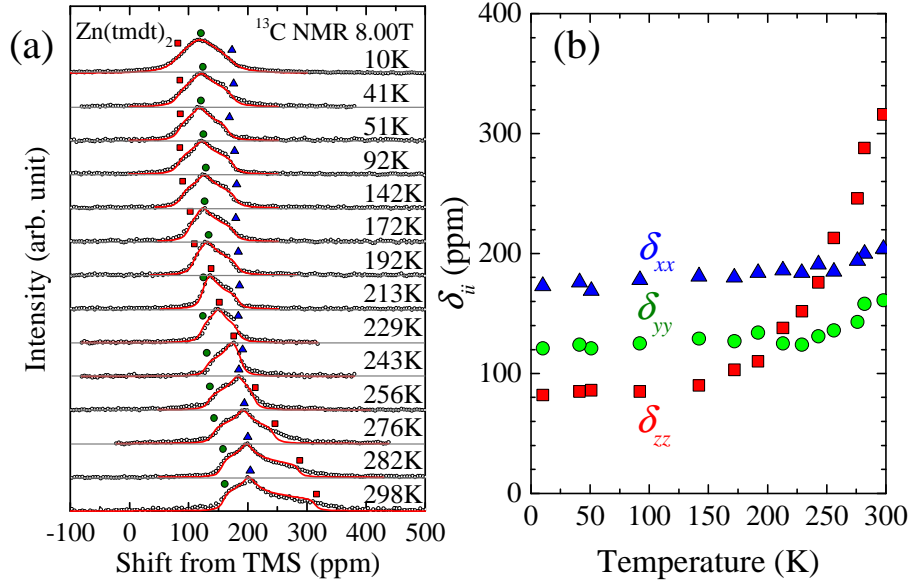}
\end{center}
\caption{(Color online) (a) Temperature dependence of $^{13}$C NMR spectra for the polycrystalline Zn(tmdt)$_{2}$. 
Solid (red) curves are fits described in the text. 
(b) Temperature dependence of the three principal values of shift tensor, ${\delta}_{ii}$ ($i=x$, $y$, and $z$). 
The symbols correspond to those indicated in (a).}
\label{Fig2}
\end{figure}

\begin{figure}
\begin{center}
\includegraphics[width=12cm,keepaspectratio]{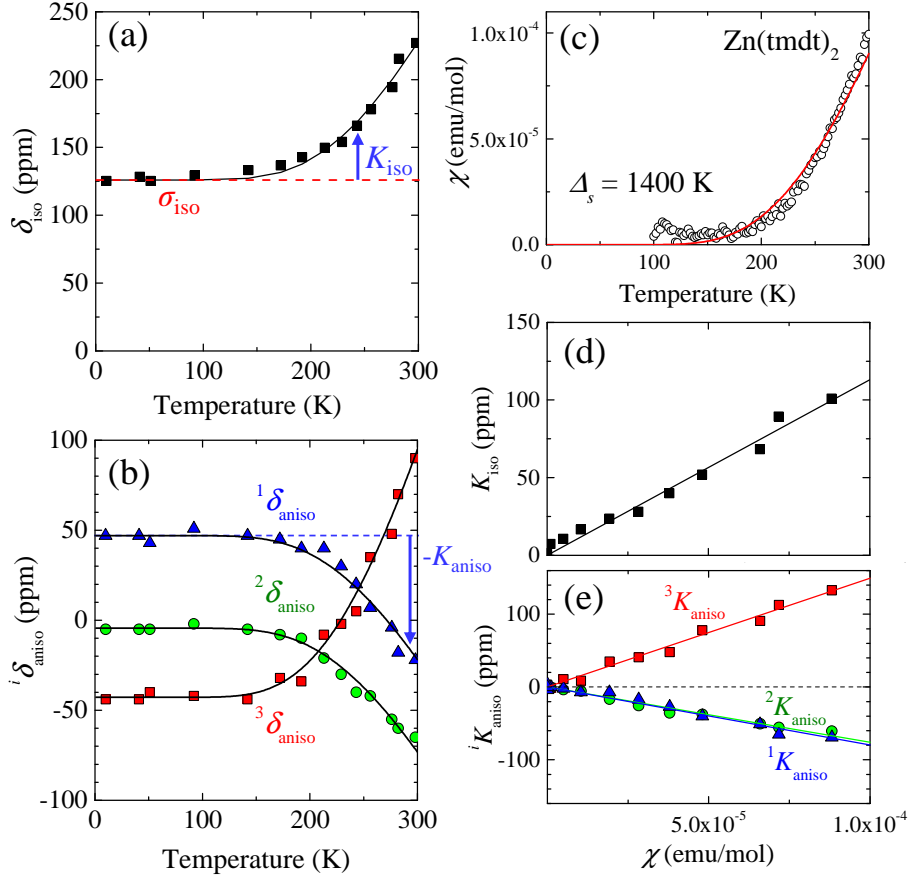}
\end{center}
\caption{(Color online)  (a) Temperature dependence of the isotropic part of NMR shift determined by the first moment of the spectra.
The red line indicates the isotropic part of chemical shift, ${\sigma}_{\rm iso}$.
(b) Temperature dependence of the anisotropic parts of NMR shifts; for example, the level of the blue line is equal to the anisotropic part of chemical shift, ${\sigma}_{1}$, and the deviation from that corresponds to $-K_{\rm aniso}$.
(c) Temperature dependence of spin susceptibility (with a Curie contribution subtracted) of the polycrystalline Zn(tmdt)$_{2}$ ,which was fitted on the basis of the singlet-triplet excitation model (red curve). 
(d) $K_{\rm iso}$-${\chi}$ and (e) $^{i}K_{\rm aniso}$-${\chi}$ plots. Solid lines are the fits of the data. 
The curves in (a) and (b) are the ${\chi}$ curve (the red curve in (c)) multiplied by respective hyperfine coupling constants determined by the slopes in (d) and (e).}
\label{Fig3}
\end{figure}

\begin{figure}
\begin{center}
\includegraphics[width=12cm,keepaspectratio]{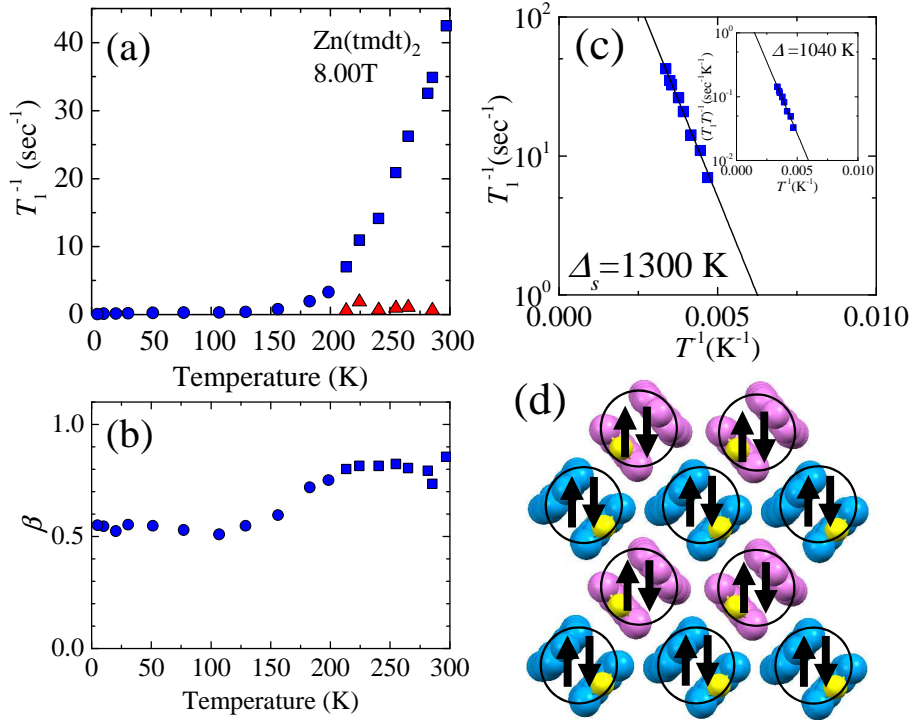}
\end{center}
\caption{(Color online) (a) Nuclear spin-lattice relaxation rate, $T_{1}^{-1}$, for Zn(tmdt)$_{2}$. 
The symbols of squares and triangles represent the relaxation rates of the sample and a minor impurity phase (see text).
(b) Temperature dependence of the stretched exponent, ${\beta}$, in the fitting of the nuclear relaxation (see text). 
(c) The activation plot of $T_{1}^{-1}$ for 200-300 K, where the relaxation rate of the sample is well separated from that of the minor impurity phase.
The inset is the activation plot of $(T_{1}T)^{-1}$. 
(d) Spin model for the $p{\pi}$ electrons in the ``${\kappa}$-type'' configuration of tmdt's in the two-dimensional layer.}
\label{Fig4}
\end{figure}

\begin{figure}
\begin{center}
\includegraphics[width=10cm,keepaspectratio]{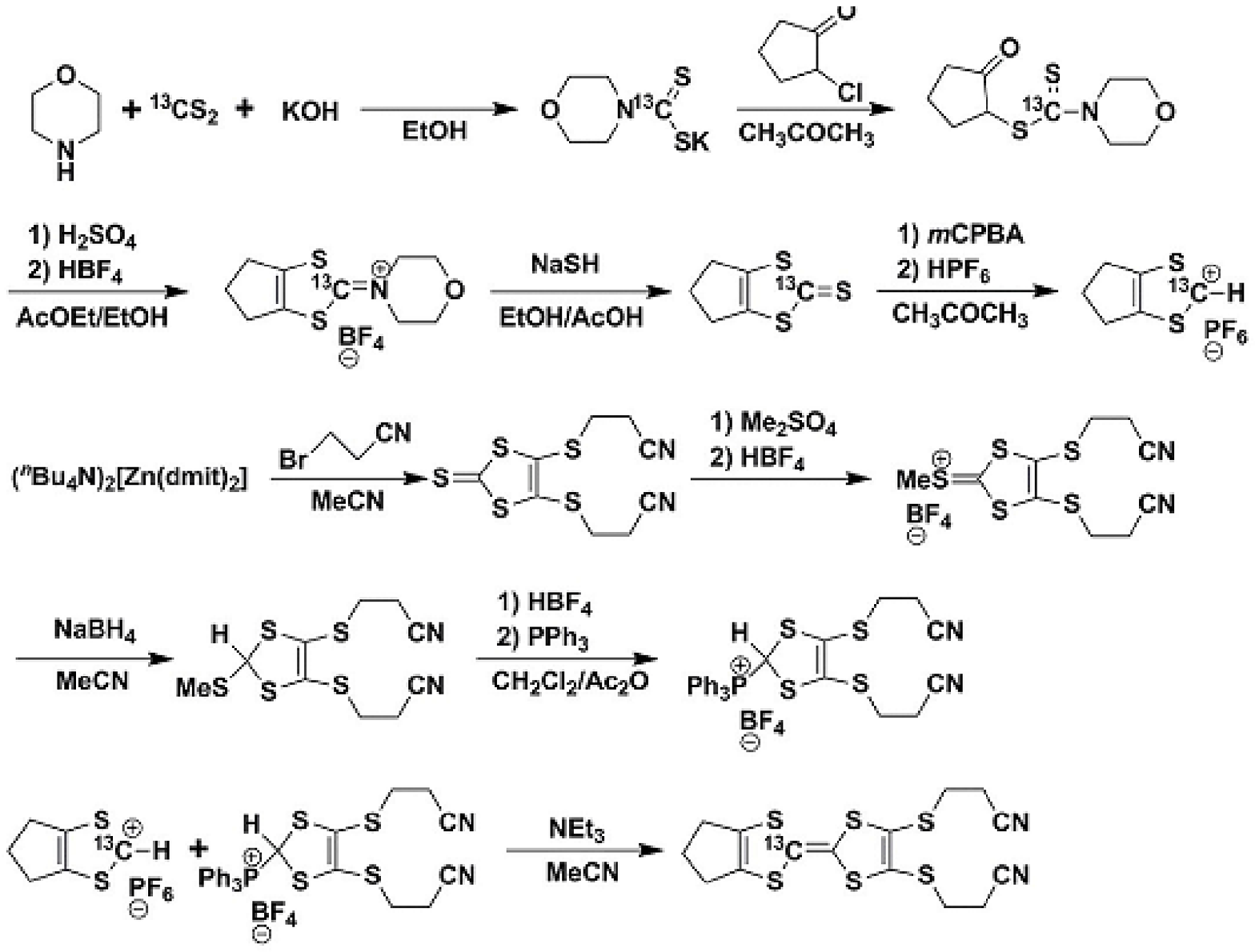}
\end{center}
\caption{Scheme 1}
\label{FigA1}
\end{figure}

\begin{figure}
\begin{center}
\includegraphics[width=10cm,keepaspectratio]{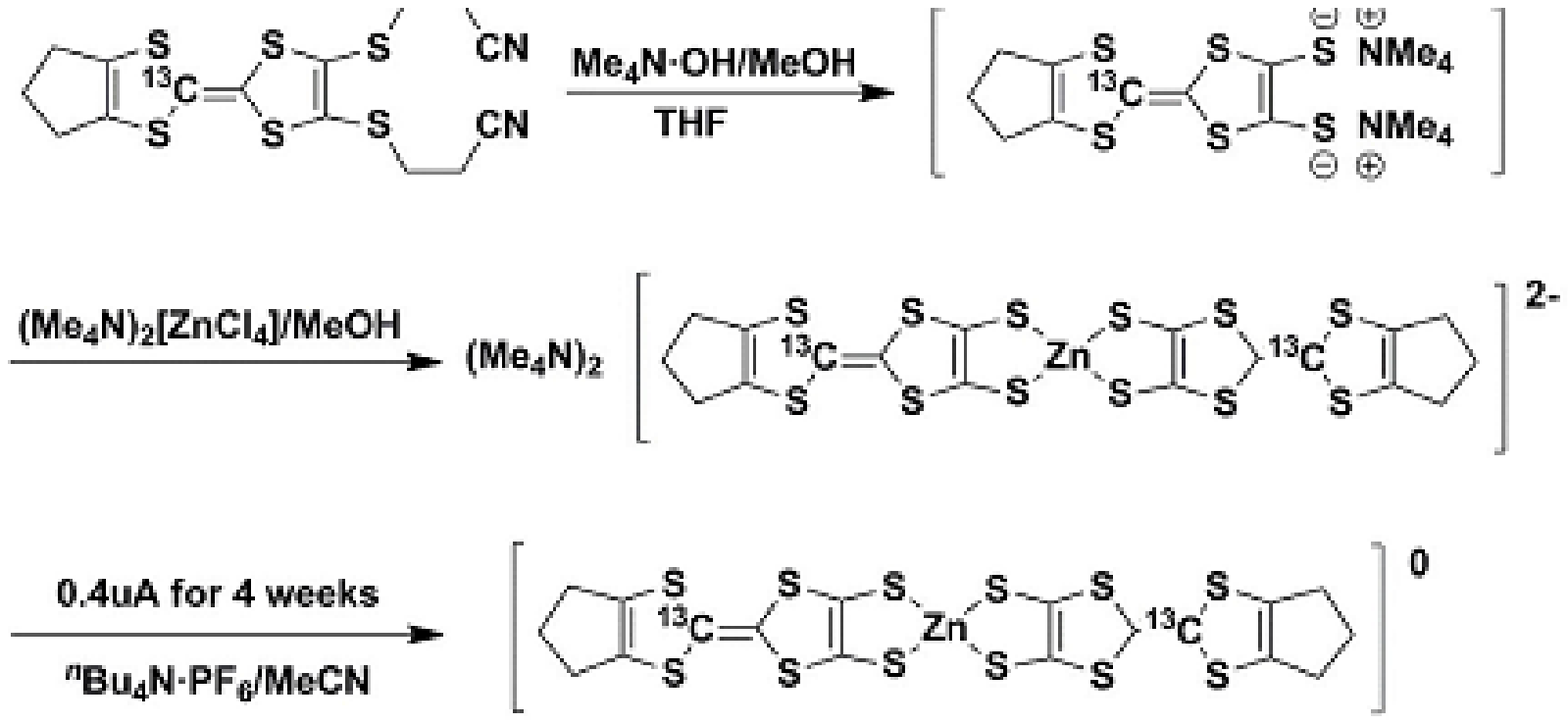}
\end{center}
\caption{Scheme 2}
\label{FigA2}
\end{figure}

\begin{table}
\begin{center}
\caption{The isotropic and anisotropic hyperfine coupling constants of $^{13}$C with the $p{\pi}$ spins in $M$(tmdt)$_{2}$.
The values of $M=$ Ni and Pt are from Ref. [13] and those of $M=$ Cu are from Ref. [9].}
\begin{tabular}{|c|c|c|c|} \hline
 &$a_{\rm iso}$ &$a_{\rm aniso}$ &$a_{\rm aniso}/a_{\rm iso}$ \\ \hline
Zn(tmdt)$_{2}$ & 3.2 & 2.1 & 0.66 \\ \hline
Ni(tmdt)$_{2}$ & 3.9 & 3.0 & 0.78 \\ \hline
Pt(tmdt)$_{2}$ & 5.1 & 4.1 & 0.81 \\ \hline
Cu(tmdt)$_{2}$ & 4.5 & 3.6 & 0.80 \\ \hline
\end{tabular}
\end{center}
\end{table}

\end{document}